\documentclass[preprint,showpacs]{revtex4}  
\usepackage{graphicx}

\begin{document}
\newcommand{\beq}{\begin{equation}}
\newcommand{\eeq}{\end{equation}}
\newcommand{\beqa}{\begin{eqnarray}}
\newcommand{\eeqa}{\end{eqnarray}}
\def\ket#1{|\,#1\,\rangle}
\def\bra#1{\langle\, #1\,|}
\def\braket#1#2{\langle\, #1\,|\,#2\,\rangle}
\def\proj#1#2{\ket{#1}\bra{#2}}
\def\expect#1{\langle\, #1\, \rangle}
\def\trialexpect#1{\expect#1_{\rm trial}}
\def\ensemblexpect#1{\expect#1_{\rm ensemble}}
\def\ol#1{\overline{#1}}
\def\kpsi{\ket{\psi}}
\def\kphi{\ket{\phi}}
\def\bpsi{\bra{\psi}}
\def\bphi{\bra{\phi}}
\def\half{\frac{1}{2}}
\def\opone{\leavevmode\hbox{\small1\kern-3.8pt\normalsize1}}

\title{Quantum and classical correlated imaging}

\author{Jan Bogdanski$^{*}$}
\author{Gunnar Bj\" ork}

\author{Anders Karlsson}
\affiliation{Department of Microelectronics and Information
Technology, Royal Institute of Technology
\\ Electrum 229, SE-164 40, Kista, Sweden\\ * correspondence: janbog@imit.kth.se}

\begin{abstract}
We outline the potential gains of quantum correlated imaging and
compare it to classical correlated imaging. As shown earlier by
A.~Gatti, E.~Bambilla, M.~Bache, and L.~A.~Lugiato,
ArXive:quant-ph/0405056, classical correlated imaging can mimic
most features of quantum imaging but at  lower signal-to-noise
ratio for a given mean photon number (or intensity). In this paper
we specifically investigate coherent correlated imaging, and show
that while it is possible to perform such imaging using a thermal
source, a coherent light-source provides a less demanding
experimental setup. We also compare the performance to what can be
obtained by using non-classical light.

\end{abstract}

\maketitle

\section{introduction}

Quantum imaging has attracted much attention in recent years
\cite{Belinskii,Strekalov,Pittman,Ribeiro,Abouraddy,AbouraddyII,Gatti,Bennink,Angelo}.
Typically, this technique exploits the quantum entanglement of the
state generated by spontaneous parametric down-conversion (PDC) in
a two-beam setup. The object to be imaged is located in one of the
beams and the information about the spatial distribution of the
object is obtained by registering the coincidence counts as a
function of the transverse position of the photon in the reference
beam, which holds a known reference object
\cite{Belinskii,Strekalov,Pittman,Ribeiro,Abouraddy,AbouraddyII}.
There has been a lively debate whether or not quantum entanglement
is a necessary ingredient to perform correlated imaging
\cite{Abouraddy, Gatti,Bennink,Angelo,Bjork}. The topic became
hotly debated after the correlated image experiment of Ref.
\cite{Pittman} was successfully reproduced using classically
correlated beams \cite{Bennink}. Subsequent analysis
\cite{Gatti,GattiII,GattiIII}, has shown that correlated imaging
can be done to some extent also by classical beams, and
experiments are underway to demonstrate correlated imaging using a
light-source with a thermal photo-count distribution. It has been
pointed out that one could obtain more information about an imaged
object if one performs coherent correlated imaging
\cite{GattiIII}. That is, instead of using only intensity
correlation, one could use the correlations between the object and
reference beams' field quadrature amplitudes, and one could then
sequentially obtain information from two (or more) non-compatible
field quadratures. We will show that, perhaps counterintuitively,
coherent correlated imaging can be done using light either from a
spontaneous parametric down-conversion source, or from a thermal
source, or better yet, using coherent light. Only the
signal-to-noise ratio scaling, with the respect to the number of
photons used, differ. The coherent light source offers practical
advantages over both other sources and gives a better performance
than a thermal source.

\section{Analysis}

In order to bring out the essence of correlated imaging, one
should carefully examine where the difference between different
imaging techniques emerges. Correlated imaging entails using two
set of modes (these sets are often referred to as beams) passing
through an object characterized by the impulse function
$h_1(x,x')$ and through a reference object, characterized by the
impulse response function $h_2(x,x')$. Computing the final
correlation signal usually amounts to evaluating a multiple
integral (or a sum) with respect to the mode functions, usually
denoted by an index $q$, cf. \cite{GattiIII}. However, in our
opinion, the multi-mode treatment is not needed in order to
understand the {\it essence} of quantum imaging.

\begin{figure}
\includegraphics[width=3.75in]{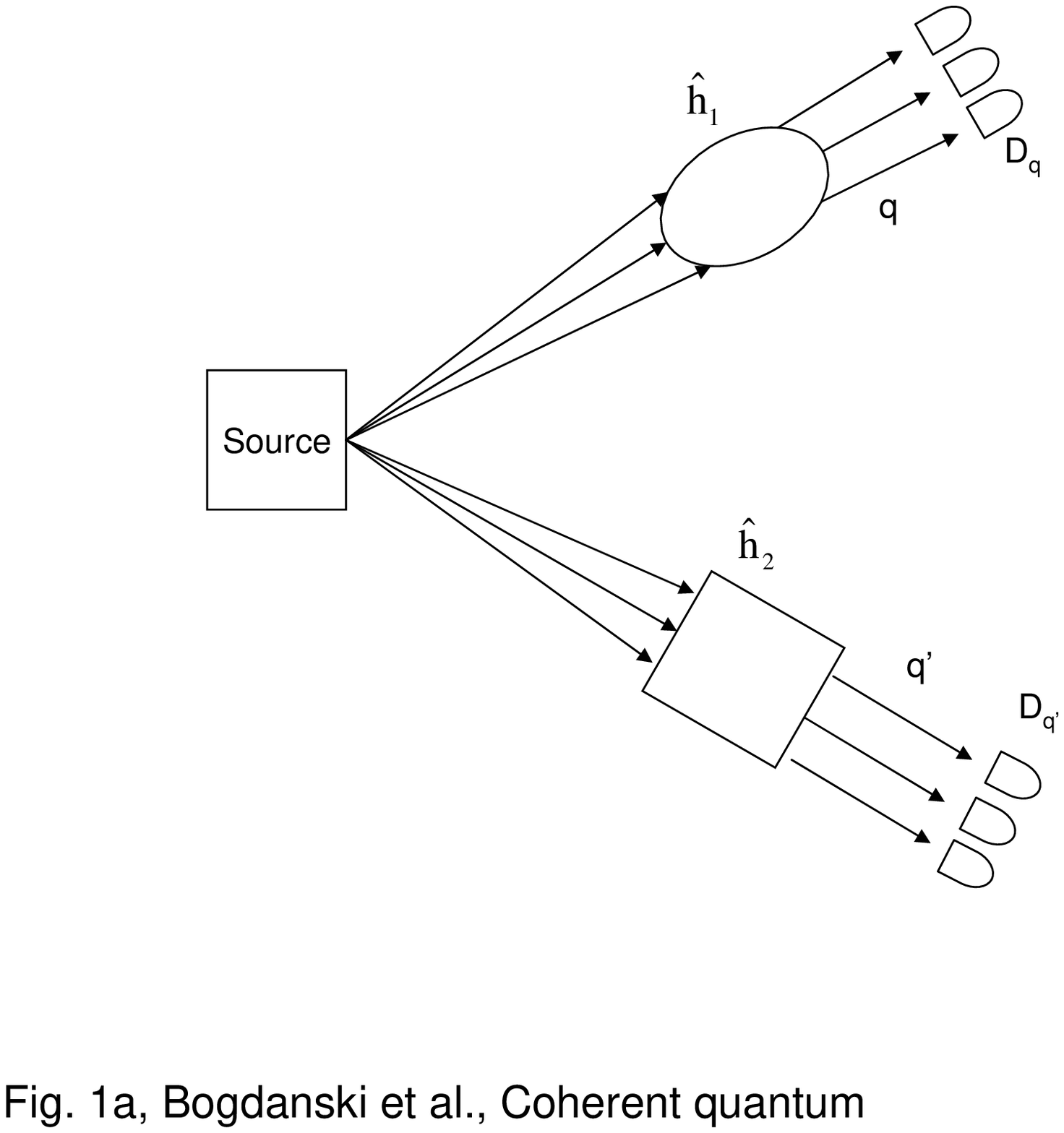}
\includegraphics[width=3.75in]{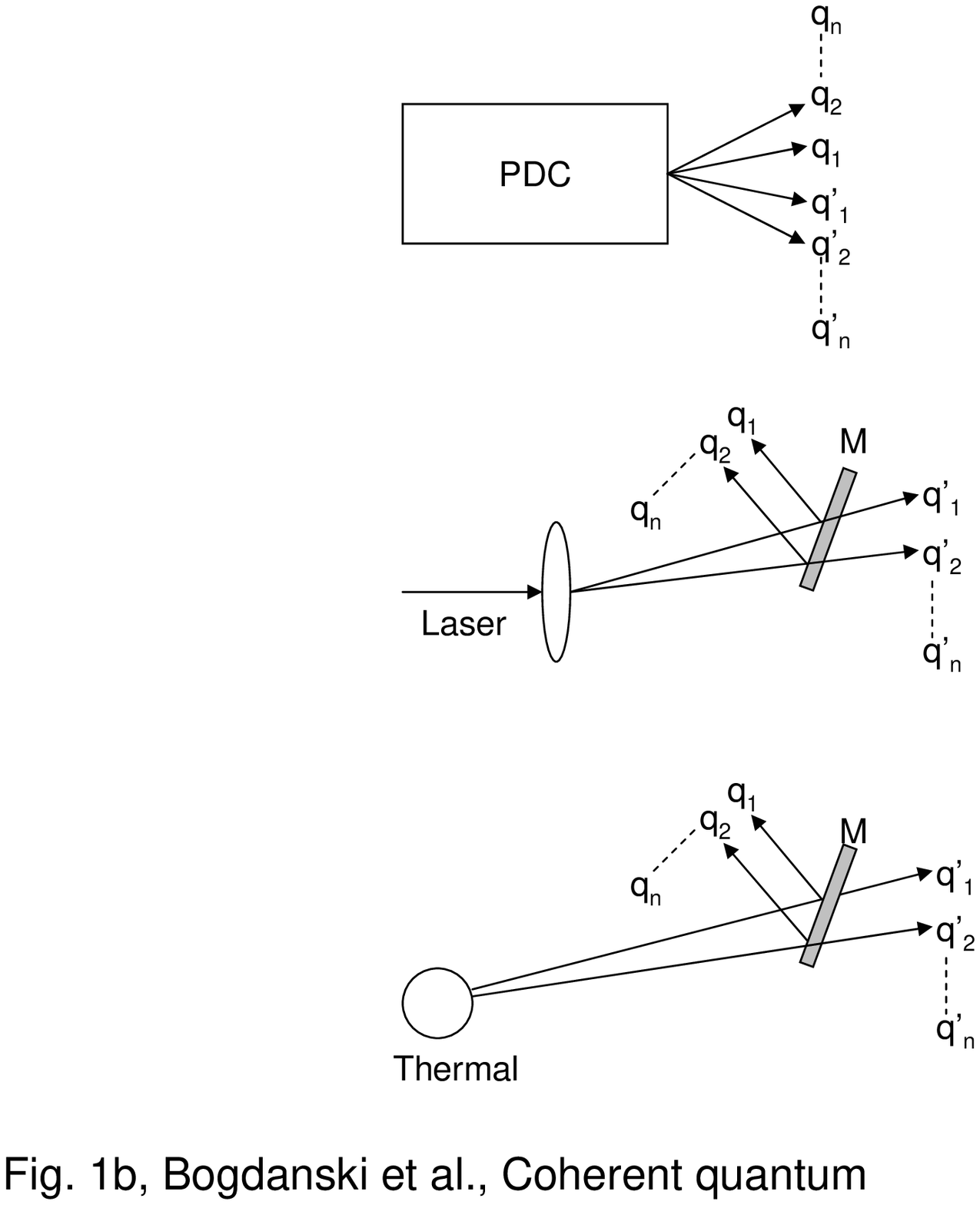}
\caption{Schematics of correlated imaging. In (a), the source with
the object and the reference depicted. In (b), the different light
sources considered in this paper are schematically illustrated. }
\label{fig: 1}
\end{figure}

In this paper we shall deal with the imaging problem by only
considering two modes (at a time). The justification for this
simplification of the problem is that, in the plane pump-wave
approximation, the transverse momentum in the parametric process
is strictly conserved, so the object (signal) $q_i$ and reference
(idler) modes $q'_i $ (in a plane-wave mode expansion),
illustrated in Fig. \ref{fig: 1}, are strictly pair-wise
correlated. Therefore, there is no first-order correlation between
different object spatial modes, nor between an object mode with
transverse wave vector $\overline{q}_i$ (with respect to the pump
wave vector) and the reference modes with wave vectors
$\overline{q}'_i \neq - \overline{q}_i$. Typically, this condition
is described by stating that the (spontaneously), parametrically
generated signal and idler beams have no transverse coherence. Of
course, it is possible to obtain any wanted transverse coherence
length (at least in principle) by appropriate filtering. However,
this is also true for any of the other two light sources
considered in this paper, and this fact does not
 change our results in any substantial way. If the PDC object and reference
beams are filtered so that they have a finite transverse
correlation length, then it is possible to find another set of
modes (than plane-wave modes) so that in this new set, the modes
are only pair-wise correlated.

We shall also assume that the detection system is arranged so that
the detection modes coincide with the set of modes where the
correlations are manifested (as common sense dictates). In an
experiment, where the pump both has a finite spatial extent
because it is focused, this entails an imaging system between the
source, the object and the (array) detector, and likewise, between
the source, reference object and its (array) detector.

In this analysis we also omit to include the impulse response
functions $h_1(x,x')$ and $h_2(x,x')$ of the object and of the
reference object. Since these are assumed to be linear, any
difference between classical and quantum imaging must stem from
the light sources, and not from the impulse response functions.
For the same reason the detectors are also omitted from the
discussion as the detectors used in the classical and the quantum
imaging are assumed to have the same characteristics.

In order to compare different sources of correlated light, the
quantity of interest in most cases is the correlation between the
intensity fluctuations (or between the field quadratures).
\begin{equation} G(x,x')= \langle I_1(x) I_2(x') \rangle -\langle I_1(x) \rangle\langle I_2(x') \rangle.\end{equation}
However, it is customary to normalize the correlation function
\cite{Wolf}, so in the following we will use
\begin{equation} C(x,x')= \frac{\langle I_1(x) I_2(x') \rangle-\langle I_1(x) \rangle\langle I_2(x') \rangle}{\sqrt{(\langle I_1(x)^2 \rangle-\langle I_1(x) \rangle^2)(\langle I_2(x')^2 \rangle}-\langle I_2(x') \rangle^2)} . \end{equation}
(Such second order correlation functions are often denoted
$g(x,x')$ in the literature.) The normalized correlation function
between field quadratures is expressed in a similar way. It is
also worth to note that the ``coordinates'' $x$ and $x'$ do not
necessarily denote spatial modes, but can refer to any set of
orthogonal modes, e.g., plane-wave modes that has no transverse
spatial dependence at all.

 To formalize the discussion, we use the
input-output relations for a linear four-port device \cite{Haus},
such as a beam splitter or a parametric amplifier. We denote the
input modes $\hat a$ and $\hat b$ and the output modes $\hat c$
and $\hat d$. The output modes correspond to any pair of modes
$q_i$ and $q'_i$ in Fig. \ref{fig: 1}. A beam splitter of (power)
transmittance $T$ then has the following operator relations:

\begin{equation} \hat c= \sqrt{T}\hat a - \sqrt{1-T}\hat b \end{equation}
and
\begin{equation} \hat d= \sqrt{T}\hat b + \sqrt{1-T}\hat a,   \end{equation}
where   $0\leq T\leq 1$. In the following we shall always assume
that $T=1/2$. An ideal parametric down-conversion amplifier obey
similar relations:
\begin{equation} \hat c= \sqrt{G}\hat a+ \sqrt{G-1}\hat b^\dagger \end{equation}
and
\begin{equation} \hat d= \sqrt{G}\hat b+ \sqrt{G-1}\hat a^\dagger ,\end{equation}
where $G$ is the (power) gain of the amplifier. The last two
equations are, in a different notation, identical to Eq. (1) in
\cite{GattiIII}. The difference in the intensity and quadrature
correlations stems from a) what type of input states are used, and
b) the appearance of the creation operator in the parametric
four-port. Let us now use these unitary transformations to compute
the correlation functions. We will consider three cases:

1) A PDC with vacuum input $\ket{  \psi_{in}}= \ket{0} \otimes
\ket{ 0}=\ket{0,0}$

2) A beam splitter with a coherent state input in one port and a
vacuum state incident on the other port: $\ket{ \psi_{in}}= \ket{
\alpha} \otimes \ket{ 0}=\ket{\alpha,0}  $

3) A beam splitter with a thermal state input on one port and a
vacuum state incident on the other port, where we use the
corresponding density operator: $\hat{\rho}_{in} =
\hat{\rho}_{therm} \otimes \ket{ 0} \bra{0}$.

The relevant fluctuation correlation for intensity correlation
imaging is
\begin{equation} C_I  = \frac{ \langle \psi_{in}|\hat c^\dag  \hat c\hat d^\dag  \hat d|\psi_{in}
\rangle-\langle \psi_{in}|\hat c^\dag  \hat c\hat |\psi_{in}
\rangle\langle \psi_{in}|\hat d^\dag  \hat d|\psi_{in} \rangle
}{([ \langle \psi_{in}|(\hat c^\dag  \hat c)^2 |\psi_{in} \rangle
- \langle \psi_{in}|\hat c^\dag  \hat c |\psi_{in} \rangle^2][
\langle \psi_{in} |(\hat d^\dag \hat d)^2 |\psi_{in} \rangle -
\langle \psi_{in} |\hat d^\dag \hat d |\psi_{in} \rangle^2])^{1/2}
}.
\end{equation}

One may also be interested in performing correlated imaging using
homodyne detection. In that case, the relevant correlations are
found in the field quadrature observables:
\begin{equation} \hat c_{i}= {1 \over 2} (\hat c+\hat c^\dagger)\end{equation}
and
\begin{equation} \hat c_{o}= {1 \over 2i} (\hat c-\hat c^\dagger) . \end{equation}
The pertinent correlation becomes (e.g., between the in-phase
quadratures):
\begin{equation} C_i  = \frac{ \langle \psi_{in}|\hat c_i \hat d_i |\psi_{in} \rangle - \langle \psi_{in}|\hat c_i  |\psi_{in} \rangle\langle \psi_{in}| \hat d_i |\psi_{in} \rangle}{( [\langle \psi_{in}|(\hat c_i )^2
|\psi_{in}\rangle - \langle \psi_{in}|(\hat c_i )
|\psi_{in}\rangle^2][ \langle \psi_{in}|(\hat d_i )^2 |\psi_{in}
\rangle - \langle \psi_{in}|(\hat d_i ) |\psi_{in}
\rangle^2])^{1/2} }
\end{equation}
Please note that, due to symmetry, the out-of-phase field
quadrature fluctuation correlation will be identical to the
in-phase fluctuation correlations for all cases considered.

Now, using any standard textbook in quantum optics, e.g.,
\cite{Loudon}, we may compute the correlations. First, we compute
the intensity and quadrature phase correlations for PDC:

\begin{equation} \hat C_{I,PDC} = 1
\label{eq: PDC CI}\end{equation} and
\begin{equation} \hat C_{i,PDC}  =\frac{{2 \sqrt {G(G - 1)} }}{{2G - 1}}
=\frac{{2\sqrt {\langle \hat n\rangle(\langle \hat n\rangle+1)}
}}{{2\langle \hat n\rangle+1}} ,
 \end{equation}
where $\langle \hat n\rangle = G-1$ is the average photon number
per mode. The result $\hat C_{I,PDC}=1$ is of course expected as
we have considered an ideal two-photon creation process.

Next, for the coherent state we obtain
\begin{equation}C_{I,coh} = 0 \label{eq: coh CI}\end{equation}
and
\begin{equation}C_{i,coh} = 0 . \end{equation}
This result is intuitive, too. As a coherent state split into two
in a beam splitter evolves into a product state of two coherent
states. Therefore, we cannot expect any fluctuation correlations.
However, please note that this does not imply that the coherent
state cannot be used for correlated imaging. It only implies a
noise limit for the imaging.

Finally, for the thermal state we get
\begin{equation}C_{I,th}  = \frac{\langle \hat{n} \rangle}{\langle \hat{n} \rangle + 1} , \end{equation}
whereas the quadrature field correlations for the split thermal
mode is
\begin{equation}C_{i,th}  = \frac{2 \langle \hat{n} \rangle}{2 \langle \hat{n} \rangle + 1},\end{equation}
where, again, $\langle \hat n\rangle$ is the average photon number
in the object and in the reference modes.

\begin{figure}
\includegraphics[width=\columnwidth]{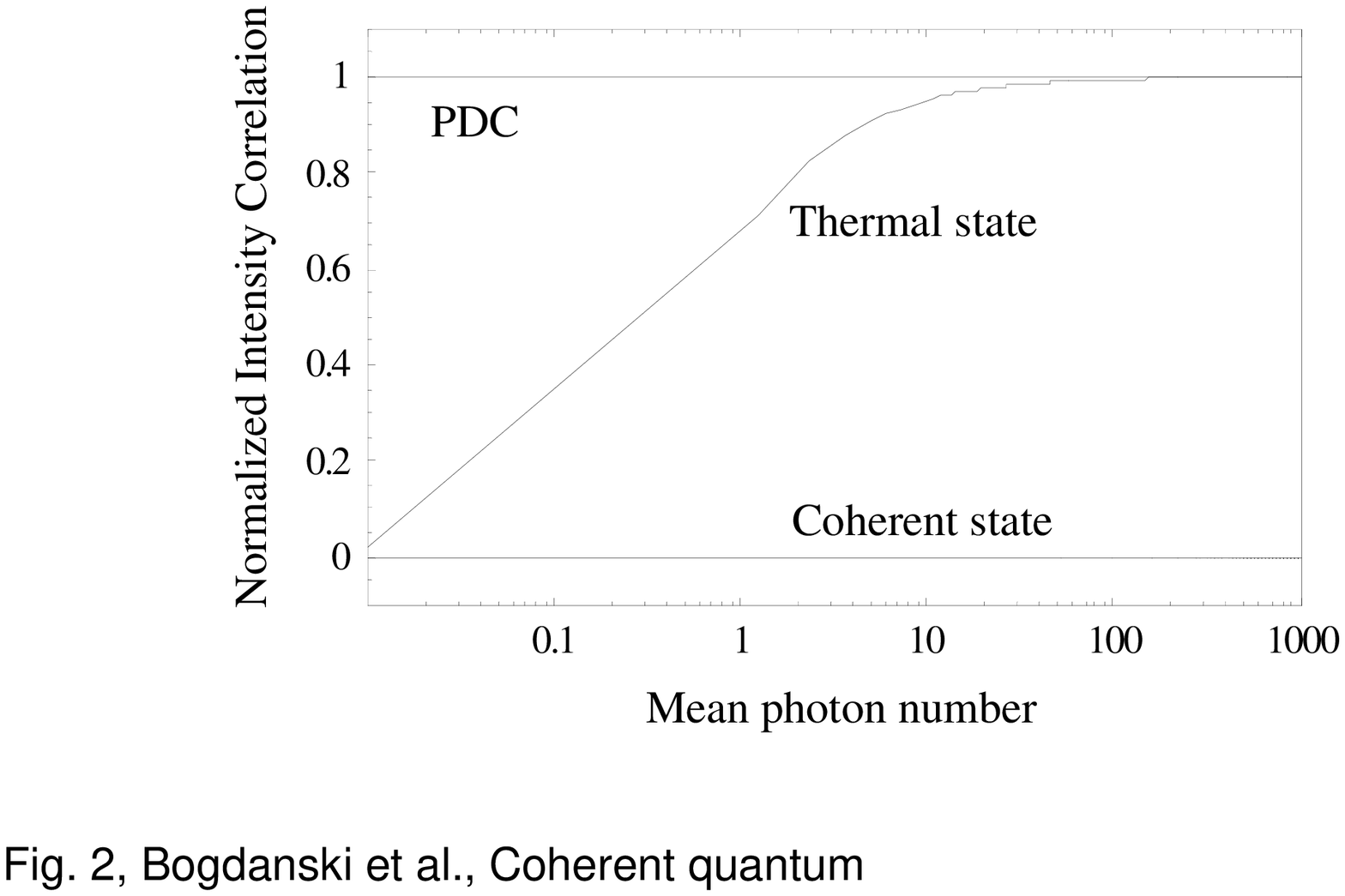}
\caption{Intensity correlation function for parametric
down-conversion (PDC), a beam splitter with a coherent state
input, and a beam splitter with a thermal state input.}
\label{fig: 2}
\end{figure}

\begin{figure}
\includegraphics[width=\columnwidth]{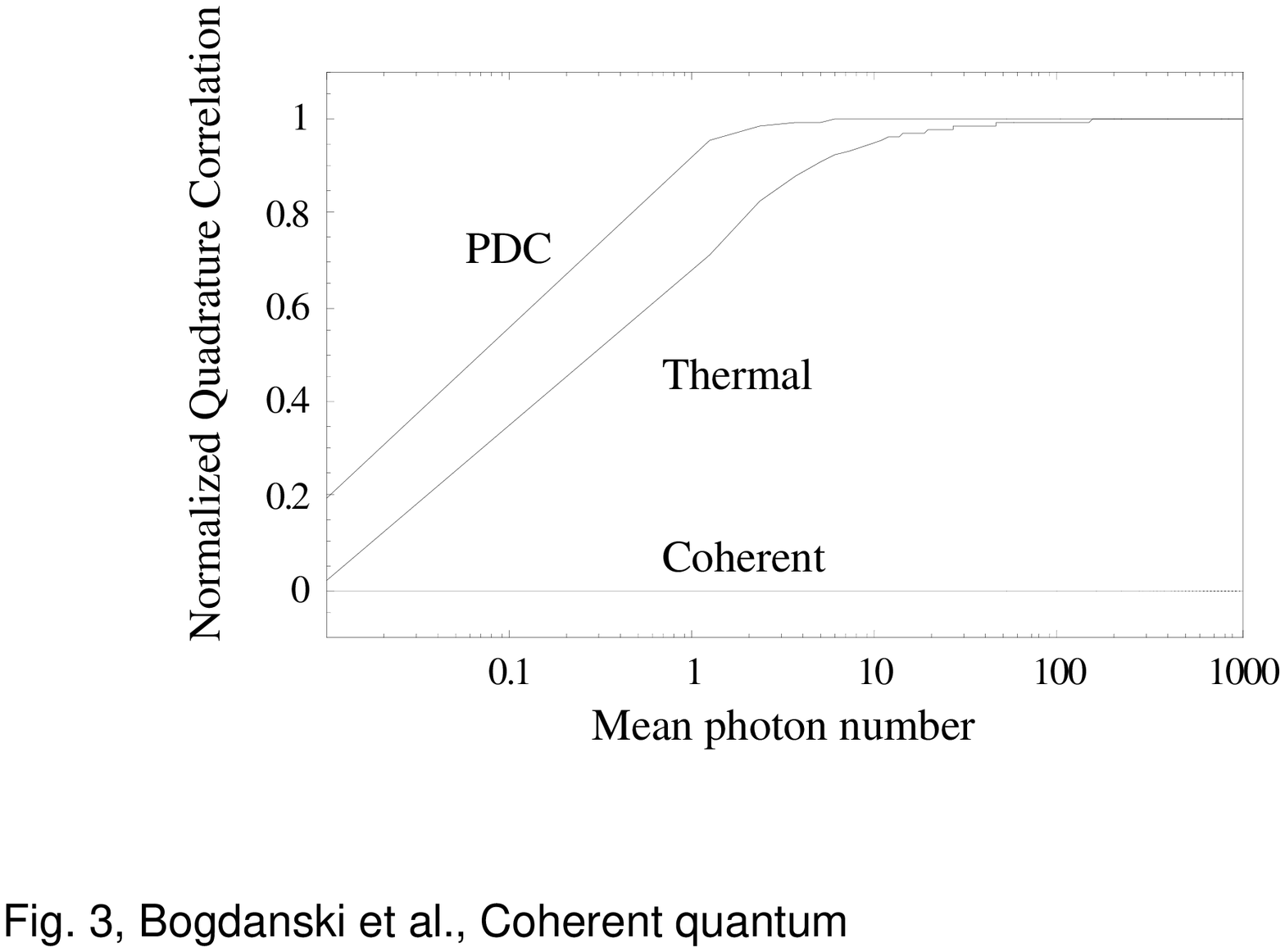}
\caption{In-phase quadrature correlation function for parametric
down-conversion (PDC), a beam splitter with a coherent state
input, and a beam splitter with a thermal state input.}
\label{fig: 3}
\end{figure}

In Fig. \ref{fig: 2}, we have plotted the intensity correlations,
and in Fig. \ref{fig: 3} the (in-phase) quadrature correlations as
a function of the photon number in the one of the two output
beams. Note that for the coherent state and the thermal state the
input average photon number is twice this number since the beam is
split in half. We see that the normalized fluctuation correlations
are almost as strong for the thermal state as for the PDC.
However, this correlation is deceiving, because the fluctuations
of a thermal state are well above the standard quantum limits, so
that the difference between the (correlated) fluctuations in the
two modes is at the standard quantum limit, as we shall see below.

In correlated imaging, the measurement limit is set by the
relative difference-fluctuations between the (in this case)
pair-wise correlated modes. If the difference in the measured
signal through the object and the reference (e.g., the difference
in transmitted photon number) is smaller that the statistical
fluctuations between the two modes, it will be difficult to detect
the difference between the object and the reference. Therefore,
the uncorrelated fluctuations set a limit to the resolution of the
correlation measurement, e.g., to how small an absorption
difference that can be detected between the object and the
reference. Because, in all considered cases we have assumed a
symmetrical generation setup with respect of the object and
reference mode ($\langle \hat{c}^\dagger \hat{c}\rangle = \langle
\hat{d}^\dagger \hat{d}\rangle$), the uncorrelated intensity
fluctuation variance is given by the expectation value of \beq
(\hat{c}^\dagger \hat{c} - \hat{d}^\dagger \hat{d})^2.\eeq The
quadrature field difference fluctuation variance is given by the
expectation value of \beq (\hat{c}_1 - \hat{d}_1)^2 .\eeq
Calculating the uncorrelated fluctuations in terms of mean photon
number per mode for the six cases we get
\begin{eqnarray} \hat V_{I,PDC} & = & 0, \label{eq: PDC VI}\\
\hat V_{i,PDC} & = & \frac{1}{2} (2 \langle \hat{n} \rangle + 1- 2
\sqrt{\langle \hat{n} \rangle( \langle \hat{n} \rangle + 1)}) , \label{eq: PDC Vi}\\
\hat V_{I,coh} & = &  2 \langle \hat{n} \rangle, \label{eq: coh VI}\\
\hat V_{i,coh} & = & \frac{1}{2}, \label{eq: coh Vi}\\
\hat V_{I,th} & = & 2 \langle \hat{n} \rangle , \quad {\rm and}
\label{eq: th VI}\\
\hat V_{i,th} & = & \frac{1}{2} . \label{eq: th Vi}\end{eqnarray}
From these equations it is clear that the PDC offers superior
performance both for correlated intensity imaging and for
correlated coherent imaging. However, in the latter case the
imaging will be beset by practical difficulties. While the field
quadratures are strongly correlated, they have a vanishing
expectation value, so that they will randomly jump between
positive and negative values (in a correlated fashion). This will
take place on a time-scale of the (first-order) coherence time of
the thermal source. That is, unless the thermal modes are very
narrowly spectrally filtered, the fluctuations will occur on a
time-scale that is to short for a typical homodyne detector to
follow.

\begin{figure}
\includegraphics[width=\columnwidth]{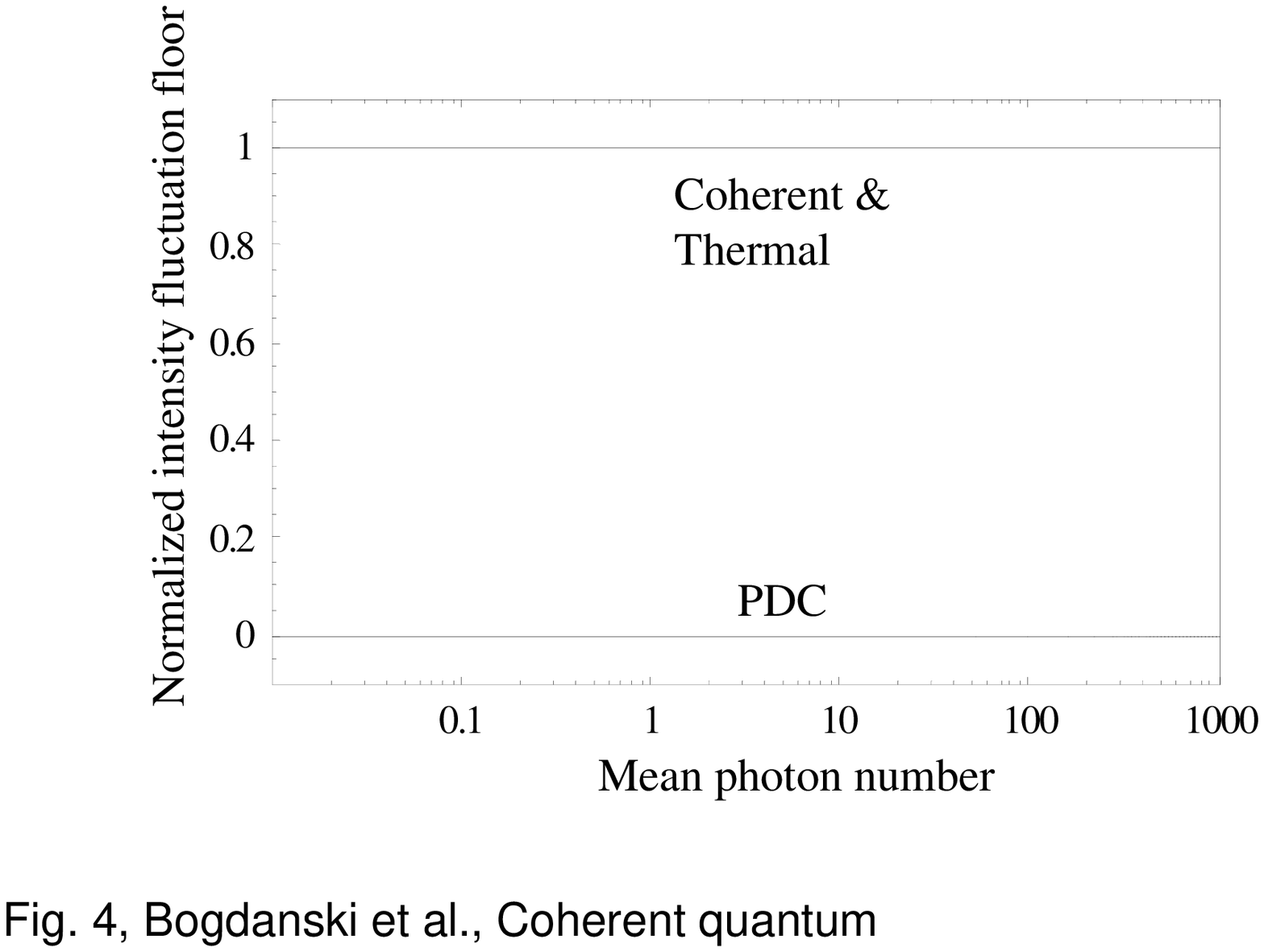}
\caption{Normalized intensity fluctuation floor for parametric
down-conversion (PDC), a beam splitter with a coherent state
input, and a beam splitter with a thermal state input. Note that
the coherent state defines the so called standard quantum limit,
abbreviated SQL.} \label{fig: 4}
\end{figure}

\begin{figure}
\includegraphics[width=\columnwidth]{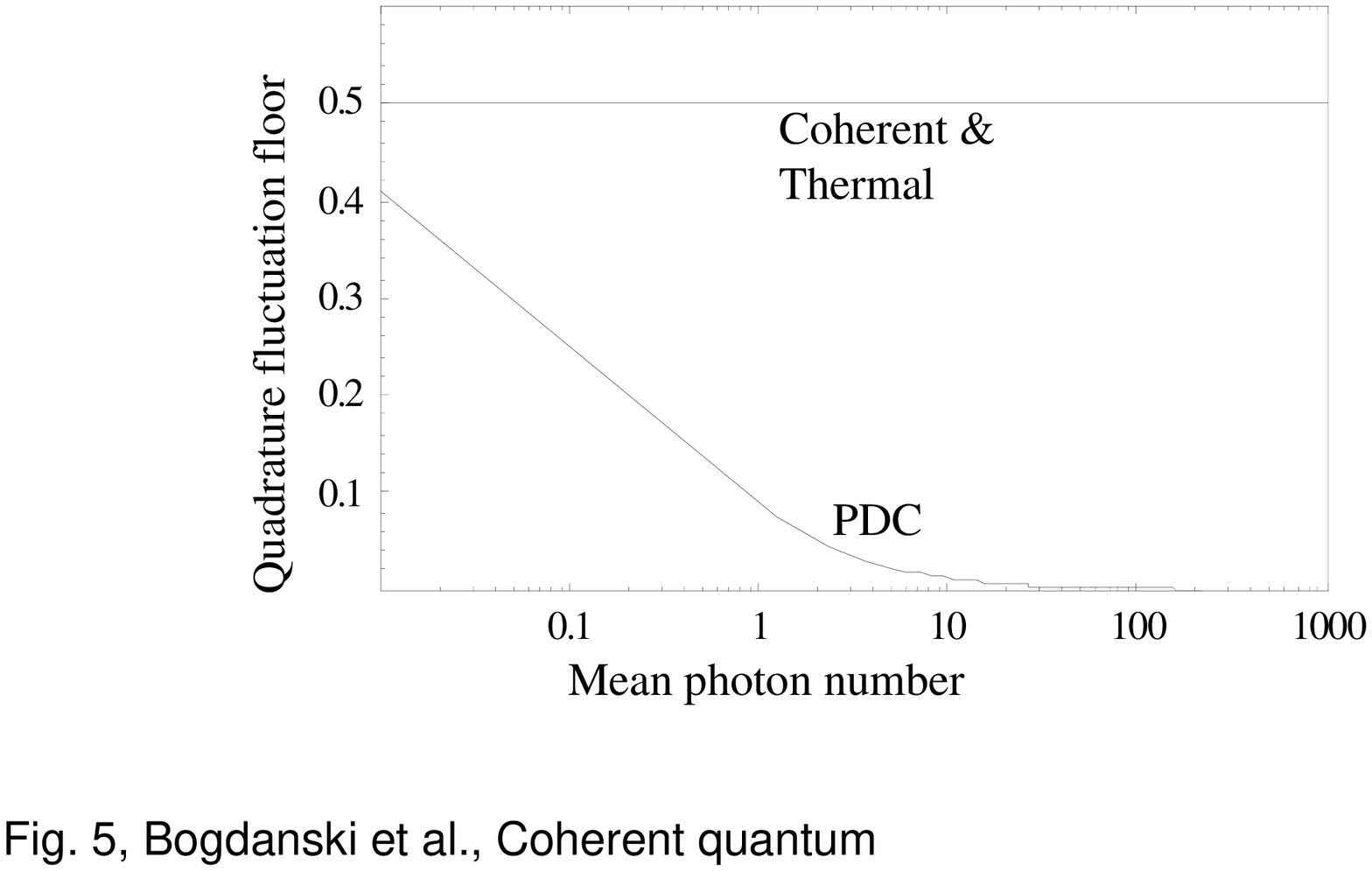}
\caption{In-phase quadrature correlation fluctuation floor for
parametric down-conversion (PDC), a beam splitter with a coherent
state input, and a beam splitter with a thermal state input. Note
that the coherent state defines the so called standard quantum
limit-SQL. } \label{fig: 5}
\end{figure}

In Fig. \ref{fig: 4}, we have plotted the uncorrelated intensity
fluctuation floor $V_{I,PDC}$, etc., normalized to the standard
quantum limit for two-mode intensity fluctuation difference, $2
\langle \hat{n} \rangle$. In Fig. \ref{fig: 5}, the uncorrelated
(in-phase) quadrature fluctuation floor is plotted. In both plots
the abscissa is chosen to be the mean photon number $\langle
\hat{n} \rangle$ in the one of the two output modes. As expected,
the coherent light will show uncorrelated fluctuations just at the
standard quantum limit, both for the intensity and for the field
quadrature fluctuations. The great advantage in this case is that
the fluctuations occur around a mean, so that using, e.g., the
same laser to produce the local oscillator needed at the detection
side of the coherent correlated imaging setup and the light used
to produce the object and reference modes, one will minimize the
problems associated with coherent imaging such as the frequency-
and the phase-stability of the local oscillators.

The thermal source, finally, will also reach an uncorrelated
fluctuation floor at the standard quantum limit. The disadvantage
from the practical point of view is that this floor is far below
the intensity fluctuation variance for each mode separately.
Hence, any small imperfection in the cancellation of the
correlated fluctuations between the object and reference mode will
lead to a large fluctuation penalty that translates to a poor
measurement resolution. Note that this is also the case for an
imperfect PDC. However, from the fundamental point of view, it is
possible to reach the same performance with a thermal source as
with a coherent source. If one aims to make coherent correlated
imaging, the field quadrature fluctuations of the thermal source
have a vanishing mean, so just like the PDC, correlated coherent
imaging with this source will present technical difficulties.

Please note that for typical photon numbers, for instance in 1 mW
of light power and one ns counting time, the photon number per
measured mode is on the order of $10^7$, implying that for all
practical purposes correlated imaging using a laser would work
perfectly! What, to date, appears to be a bit unclear within the
quantum physics community is that correlated imaging measuring the
electromagnetic quadratures also works well with classical
coherent state sources, and even, in principle, for a thermal
source. From an engineering viewpoint this is perhaps less
surprising.

\section{Discussion}

Given these results, most already pointed out in
\cite{Gatti,Bennink,Angelo,GattiII,GattiIII}, we see that if one
only measures one observable at a time, e.g., we image an object
and record the object and the reference mode intensities,
everything quantum can also be done classically, using a coherent
state generated by a laser or a thermal light source. The price to
be paid is lower signal-to-noise ratio for a fixed detected photon
number. In terms of engineering, the classical correlated imaging
can be done with less effort, and in terms of {\it overall} energy
efficiency (taking into account the low pumping efficiency in
non-linear optical processes) with fewer photons than working with
a parametric source. For such cases, we argue that the
justification for using quantum imaging can be questioned. Gatti
{\it et al.} argue that the imaging can be done with different
wavelengths in the object and in the reference mode, which is
undoubtedly true. This is one advantage we can see. [However, it
is also possible to do intensity correlation imaging using two
coherent states (lasers) operation at different wavelengths. Their
difference fluctuations will of course be uncorrelated, but this
is already the case when the two beams originate from the same
laser as shown by Eq. (\ref{eq: coh CI}).] The other potential
advantage with quantum correlated intensity imaging is seen in the
few photon regime, where the strong quantum correlations offer a
clear advantage, c.f. \cite{GattiIII}, see also \cite{Belinsky}.
This correlation has been proposed to be used in military ranging
applications \cite{Rarity}.

The main new result we report is that nothing is gained by using a
thermal source rather than a coherent source when doing classical
correlated imaging. One may be lead to believe otherwise by
looking at the pairwise correlations between modes in a split
thermal beam. The correlations are almost as strong as the
correlations between parametrically generated signal and idler
beams. However, this strong correlation is deceiving, because the
fluctuations in each of the thermal modes are much higher for a
thermal source than for an equally intense coherent source. The
imaging signal-to-noise ratio for the thermal and coherent sources
are the same, but the coherent source will make an experimental
implementation simpler, in particular if coherent imaging is
employed.

What then is the essence of correlated quantum imaging?  In our
opinion, this is to be found in the nature of entanglement, and
how much this is exploited. For instance, as discussed in section
IV of \cite{GattiIII}, and realized experimentally in
\cite{Dopfer,Howell}, when modifying the experiment in order to
utilize the correlations between the different observables, one
also needs to modify the classical states needed to mimic the
quantum statistics. That is, a classical state can have strong
correlations between some pair of observables, but unlike a
quantum state it cannot simultaneously have strong correlations in
a complementary pair of observables \cite{Dopfer,Howell}. This is
the essence of all Bell-inequalities, where a quantum state's
summed correlations between incompatible observables exceed the
limit set for any locally realistic theory, such as classical
physics. Another case of interest is if one images true quantum
objects, i.e., one ``captures'' more than a single observable from
the object. In this meaning, quantum imaging becomes very much
related to (multi-mode) quantum teleportation, a relationship that
could be worthwhile to explore further. However, as soon as the
``capture'' is collapsed by a readout of some observable, we are
back to the imaging described above.

\section{Acknowledgements}

This work was supported by the Swedish Defence Material
Administration (FMV), the Swedish Foundation for Strategic
Research (SSF), and the Swedish Research Council (VR). One of the
authors (GB) would like to acknowledge stimulating discussions
with Professor L.~A.~Lugiato.

\pagebreak

\end{document}